\documentclass[twoside,slac_one]{revtex4}
\usepackage{graphicx}
\usepackage{fancyhdr}
\usepackage{amsmath} 
\usepackage{bm}
\usepackage{amsxtra}
\usepackage{amssymb}
\usepackage{amsthm}
\usepackage{latexsym}
\usepackage{lscape}

\newcommand{\MET}{{\slash\!\!\!\!E_T}}
\newcommand{\DZero}{D$\slash\!\!\!0$}
\newcommand{\fbi}{\mathrm{fb}^{-1}}

\begin{document}

\title{Reconciling the CDF $Wjj$ and single-top-quark anomalies}

\author{Zack Sullivan}
\affiliation{Department of Physics, Illinois Institute of Technology, Chicago, Illinois 60616-3793, USA}
\author{Arjun Menon}
\affiliation{Department of Physics, Illinois Institute of Technology, Chicago, Illinois 60616-3793, USA}

\begin{abstract}
  We demonstrate that there is no evidence of any $Wjj$ excess or
  deficit within CDF data if a data-derived background estimation
  that includes single-top-quark production is used instead of a Monte
  Carlo estimate.  Instead, when coupled with the CDF measurement of
  single-top-quark production, a more interesting anomaly exists
  within CDF data: namely, there are too many $W+0$ $b$-tag and $W+2$
  $b$-tag events, and too few $W+1$ $b$-tag events.  As we previously
  predicted, there is no significant evidence of any of these
  anomalies in the \DZero\ data set.
\end{abstract}

\maketitle

\thispagestyle{fancy}

\section{Introduction}

Great excitement was caused by the observation by the CDF
Collaboration of a large apparent peak in $Wjj$ that was suggested in
Ref.~\cite{Aaltonen:2011mk} to be the result of a Gaussian peak
sitting on top of the $Wjj$ continuum background.  In
Ref.~\cite{Sullivan:2011hu}, we demonstrated how an existing excess in
the CDF measurement of single-top-quark production would translate
into the $Wjj$ signal region, and fit the apparent excess, assuming it
was a fluctuation of single-top-quark production.  We concentrated in
that paper on single-top-quark production, rather than $t\bar t$
production, because of a statement in the CDF paper that the $t\bar t$
contribution had been fit to data, but single-top had only been
modeled by Monte Carlo.

This proceedings constitutes a significant update to Ref.\
\cite{Sullivan:2011hu} that clearly demonstrates the excesses in $Wjj$
and in the single-top measurements are the same.  Unlike our paper, we
are now able to show that no corrections to single-top are required to
fit the data.  What is less clear is whether the events are physically
due to single-top, $t\bar t$, or most likely a combination of many
things.  While we do not solve all problems, we demonstrate these
anomalies reside solely within the CDF data, and reclassify the
issues as a result of an anomaly in the number of $b$-tagged jets in
the $Wjj$ sample.

In order to clearly identify where the real discrepancies exist, we
first reexamine the data to learn more about the $Wjj$ anomaly, and
then show its resolution.  In Section \ref{sec:jets}, we point out
that the CDF data set strongly suggests that the excess is partly due
to a feed down effect of the $W+3$-jet sample into the $W+2$-jet
sample.  The heart of this update is Section \ref{sec:fits}, where we
refit the CDF data with single-top as extracted from data
\cite{Aaltonen:2010jr}, and compare to the fit in
Ref.~\cite{Aaltonen:2011mk}.  Here we explore several problems with
the original CDF fit that all disappear under the assumption the
excess is pure single-top.  We also demonstrate that an identical
prediction for \DZero\ perfectly fits their data set.  Finally, in
Section \ref{sec:what} we conclude by addressing the question again of
whether the apparent excess is single top or something else.

\section{$W+3$-jet feed down to $W+2$-jets}
\label{sec:jets}

\begin{figure}[htb]
\includegraphics[width=3.65625in]{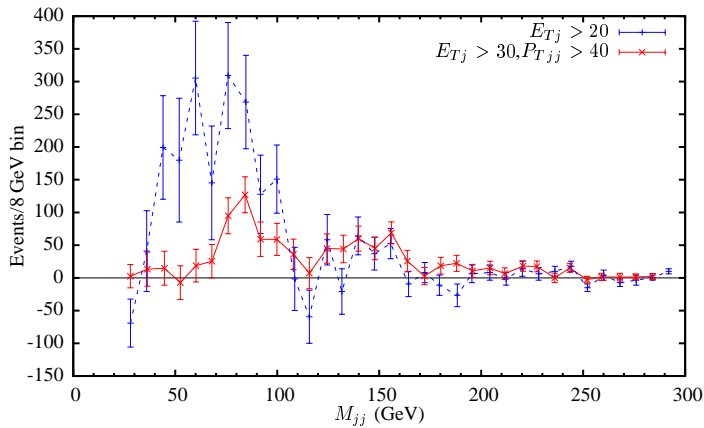}
\caption{$M_{jj}$ for $Wjj$, where jets are defined to have
  $E_{Tj}>30$ GeV (solid red) and $E_{Tj}>20$ GeV (dashed
  blue). \label{fig:lowercut} }
\end{figure}

As additional tests have been performed on the CDF data set to try to
pin down the source of the $Wjj$ excess, we have attempted to place
these results in context.  In particular, one measurement by CDF
provides an essential clue as to the source of the excess.  In Fig.\
\ref{fig:lowercut} we overlay two data sets on top of each other: the
original $Wjj$ sample from the CDF paper, and the CDF $Wjj$ sample
where the definition of a jet was loosened to $E_{Tj}>20$ GeV from the
original $E_{Tj}>30$ GeV \cite{cdfweb}.  In addition, the $P_{T\,jj}$
cut is removed in the looser set.

When first looking at this figure, it seems that perhaps there is a
mistake in the data.  The bins with more tightly defined jets have
systematically \textit{more} events than the bins with more loosely
defined jets above 100 GeV.  If the looser jets were a super-sample,
this would not be possible, and this would indicate a clear
inconsistency.  After some consideration, however, it is apparent the
looser defined jets are only a \textit{partial} super-set of the
tighter jets.  In particular, because this is an exclusive 2-jet final
state, by lowering the threshold for jet acceptance, several 3-jet
events that had a jet between 20--30 GeV are thrown away.

The conclusion is that, assuming this data is correct, a significant
portion of the excess appears to be coming from proto 3-jet events
that are sneaking in to the 2-jet sample because of the effectively
weaker jet veto.  This is consistent with the CDF check that a more
inclusive sample (allowing in 3 jets) did not change the result,
because the relevant 3-jet events were already there.  As we'll see
below, this is what we see in single-top as well.

\section{Refitting the CDF data}
\label{sec:fits}

In order to track down what could have faked a peak in the 120--160
GeV, we investigated the influence of several effects: jet energy
resolution, sensitivity to particular cuts, etc.  In our initial paper
\cite{Sullivan:2011hu} we thought we might need strong assumptions
about these effects to explain what was going on.  It has turned out
we do not.  After more thorough investigation we find that
\textit{absolutely no corrections} are required to explain the excess
as anything but single-top.  Hence, we hold off until Sec.\
\ref{sec:what} any mention of the role of jet energy resolution
effects, etc., as we now find they are not needed.

Using shapes from pure vanilla MadEvent, and normalizations from the
CDF fit to single-top, we can perfectly explain not only the excess
between 120--160, but in fact the entire region 28--300 GeV.  We find
our fits solve several problems with the initial CDF fits that were
not emphasized before, but indicate a more complex story than just an
excess in the 120--160 GeV region.  In Subsec.\ \ref{ssec:cdffit} we
reexamine the CDF sample and point out some statistical problems with
the CDF fit.  We then add our estimate of single-top in Subsec.\
\ref{ssec:stfit}, and demonstrate the problems vanish.  Finally, we
show our fit to \DZero\ in Subsec.\ \ref{ssec:dzero}.

\subsection{Re-examining the CDF fit}
\label{ssec:cdffit}

\begin{figure}[htb]
\includegraphics[width=3.75in]{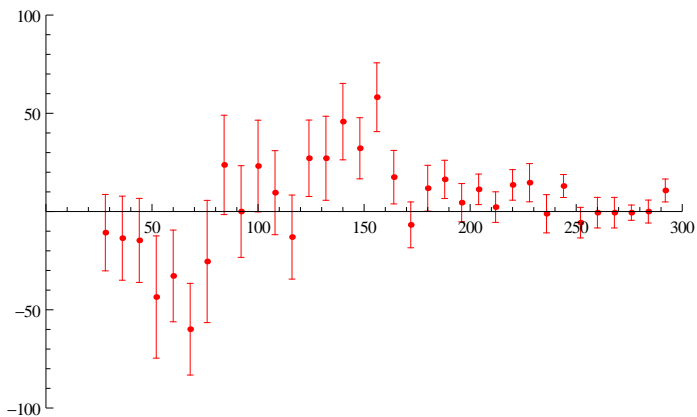}
\caption{$M_{jj}$ for remainder after subtracting all backgrounds from
the data. \label{fig:cdfsubbase} }
\end{figure}

In order to make a quantitative statement about how top explains the
excess, we start by extracting the data from
Ref.~\cite{Aaltonen:2011mk} for three sources: the measured data, the
$WW/WZ$ diboson peak, and the rest of the background fit (which
includes everything else: $Wjj$, $t\bar t$ fit to data, single-top
from Monte Carlo, and other smaller components).  We then proceed to
look at the remainder of the data after subtracting all backgrounds
from the measurement.  The result appears in Fig.\
\ref{fig:cdfsubbase}.

What is readily apparent from Fig.\ \ref{fig:cdfsubbase} (more so than
Fig.\ 1 of the CDF paper which leaves in dibosons) is that there is a
systematic problem with the $M_{jj}$ shape across the entire spectrum,
from 28--300 GeV.  In particular, the \textit{deficit} of events below
$WW/WZ$ threshold is more worrisome than the excess between 120--160
GeV.  However, we see that the excess in fact exists everywhere from
84--300 GeV.  Already we can perceive that we will want a broad
kinematic distribution to fix this, but that is discussed in the next
Subsection.

\begin{figure}[htb]
\includegraphics[width=3.75in]{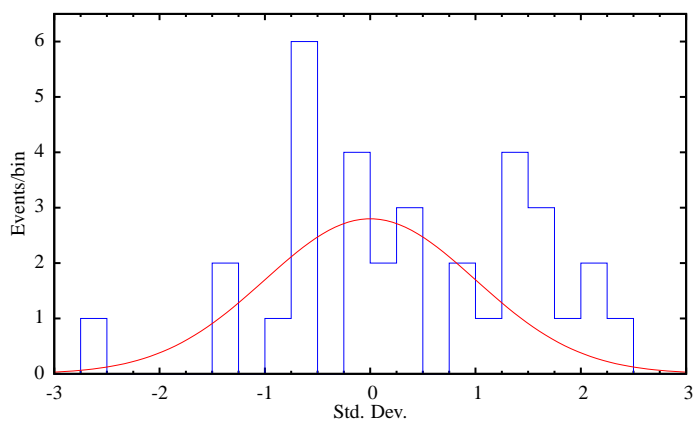}
\caption{Distribution of error residuals in CDF fit compared to a
  Gaussian distribution. \label{fig:cdfstats} }
\end{figure}

To explore this a bit more, we perform the $\chi^2$ fit on the data we
extract to quantify the quality of the background fit.  When fitting
over 28--200 GeV we get a $\chi^2/\mathrm{d.o.f.} = 44.5/19$.  It is
not surprising the fit is poor, as it is already clear from Fig.\
\ref{fig:cdfsubbase} that the data does not follow a Gaussian
statistical distribution.  We emphasize this point in Fig.\
\ref{fig:cdfstats}, where we plot the distribution of residual errors
and compare it to a Gaussian.  The data exhibits significant positive
skew and kurtosis --- i.e., there is a missing systematic effect.  The
CDF paper attempted to solve this by adding a Gaussian peak and
refloating the background and signal fits.  A simple count of the
number of points above and below the fit with a Gaussian still shows a
massive skew (14/30 points are at least $1\sigma$ above the fit) ---
hence, it will not be a good fit either.  The need for exotic partial
corrections disappears below, so we will not dwell on the Gaussian
hypothesis.

In conclusion, it is clear that the backgrounds considered in the
analysis do not fit the data, despite their normalizations being
floated.  As we are about to see, this is solved by adding single-top
as extracted from data.

\subsection{Adding data derived single-top}
\label{ssec:stfit}

\begin{figure}[htb]
\includegraphics[width=3.65625in]{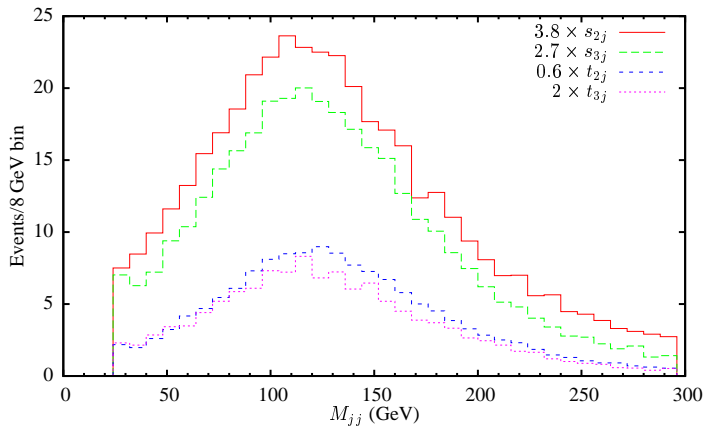}
\caption{$s$- and $t$-channel contributions to $Wjj$ after all cuts
  from initial 2-jet and 3-jet samples multiplied by $K$-factors
  extracted from CDF single-top data. \label{fig:mjjmodes} }
\end{figure}

In order to quantify the effect of translating the CDF single-top
measurement into the $Wjj$ channel, we do the following: We run
$s$-channel and $t$-channel single-top in MadEvent, producing both
$Wjj$ and $Wjjj$ final states.  $Wjjj$ is an NLO correction to LO, but
was separately extracted by CDF in their single-top fit.  We apply all
of the same cuts as the CDF paper, and produce $M_{jj}$ for each
sample.  In the case of $Wjjj$, many events are killed by the jet
veto, but not all.  In Fig.\ \ref{fig:mjjmodes} we show the shapes for
each channel, based on the number of jets initially produced.  Note
the shapes are all about the same.  In fact, under unit normalization,
the shapes are nearly identical, with 2-jet $s$-channel slightly
harder above 200 GeV, and 2-jet $t$-channel slightly softer below 80
GeV.

The distributions in Fig.\ \ref{fig:mjjmodes} are normalized to
Ref.\ \cite{Aaltonen:2010jr}, and the supporting web page, which
\textit{did not} find the predicted Standard Model-like ratio of these
modes.  In particular, $t$-channel production was extracted in the CDF
analysis as being $0.6\times$ the expected size in 2-jets, and
$2\times$ in 3-jets, and $s$-channel production was found to be too
large by a factor of $3.8$ in 2-jets, and $2.7$ in 3-jets.  An
essential point is that this is the same raw data set that $Wjj$ was
drawn from --- the same trigger, and almost the same integrated
luminosity.

We want to see what the CDF single-top data set predicts for $Wjj$, so we
do the following:
\begin{enumerate}
\item We first remove a Standard Model size prediction for single-top
  from the CDF background estimate we extract from their paper (call
  this $Wjj_r$ for residual).  We use NLO $K$-factors (which have been
  checked in the past after cuts) of 1 for $t$-channel and $1.5$ for
  $s$-channel \cite{Sullivan:2004ie}.
\item We refit the data using a minimal $\chi^2$ test on three samples:
\begin{enumerate}
\item $a\times Wjj_r$ --- all backgrounds as predicted by CDF except for
  dibosons and single-top.
\item $b\times VV$ --- $WW/WZ$ dibosons.
\item $c\times$single-top --- where we add $0.6\times t_2 + 2\times t_3 +
3.8\times s_2 + 2.7\times s_3$ to match the ratios extracted by CDF.
\end{enumerate}
\item Compare $a$, $b$, $c$ to 1, and to the CDF fit.
\end{enumerate}

Before we show the results, a word on these fits.  While we have
chosen to use the ratio extracted from CDF, we immediately notice that
the deficit in $t$-channel 2-jet is largely cancelled by the excess in
$t$-channel 3-jet, leaving us mostly sensitive to additional
$s$-channel.  However, as the shapes are all about the same, and the
quoted CDF single-top errors are large (50--80\%), it is easily
conceivable that there could be a different mixture in the sample.  We
choose the ratio observed by CDF to perform our fit, but in the end,
we are really just fitting a shape.  It turns out the normalization is
also perfectly consistent.

\begin{figure}[htb]
\includegraphics[width=3.65625in]{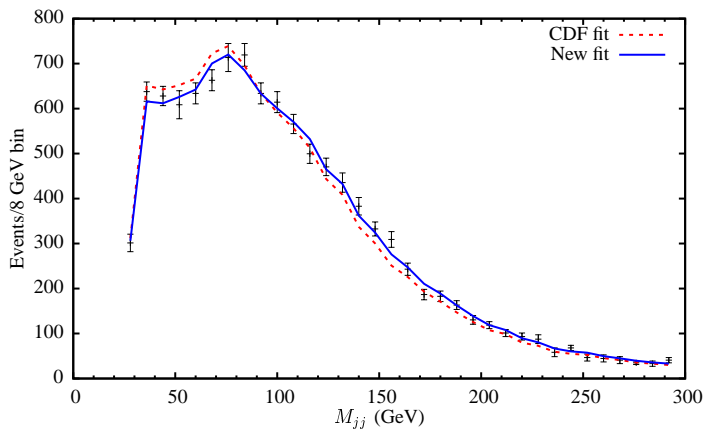}
\caption{Comparison of $Wjj$ data with CDF fit (red dashed), and fit
  with single-top extracted from CDF data (blue
  solid). \label{fig:datavfits} }
\end{figure}

In Fig.\ \ref{fig:datavfits} we compare the CDF fit to the data with
the the new best fit using $1.4\times$ single-top as extracted from
data.  It is clear that the fit including single-top is now consistent
everywhere from 28--300 GeV.  Our fit of the CDF
$\chi^2/\mathrm{d.o.f.}$ of $44.5/19$ improves to
$\chi^2_\mathrm{new}/\mathrm{d.o.f.} = 25.7/26$ (we fit 30 points with
3 variables).  Our best fit finds we need $0.91\times$ as much $Wjj_r$
and $WW$ as was required in the CDF fit, but as those normalizations
were floated in the CDF fit as well, it is not a surprise.

\begin{figure}[htb]
\includegraphics[width=3.75in]{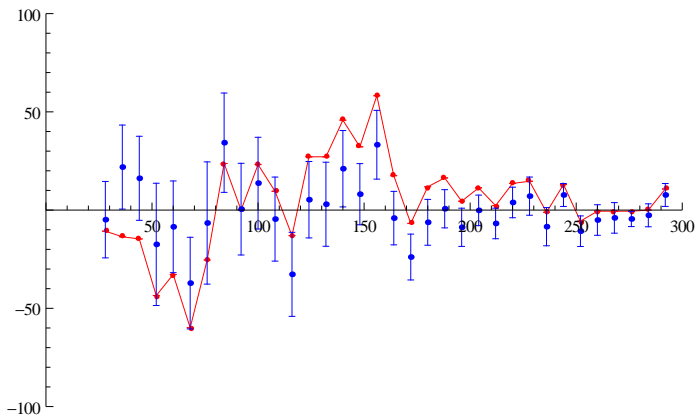}
\caption{Comparison of data minus background with the CDF fit (red
  line), and the new fit using single-top extracted from CDF data
  (blue error bars). \label{fig:subnew} }
\end{figure}

To stress that the residual is statistically removed, we compare the
background subtracted data before and after the new fit in Fig.\
\ref{fig:subnew}.  The red line represents the central values of the
original baseline CDF subtracted data, and the blue error bars
represent what remains after subtracting the new background fit.  The
systematic deficit(excess) below(above) $WW$ is gone.

\begin{figure}[htb]
\includegraphics[width=3.75in]{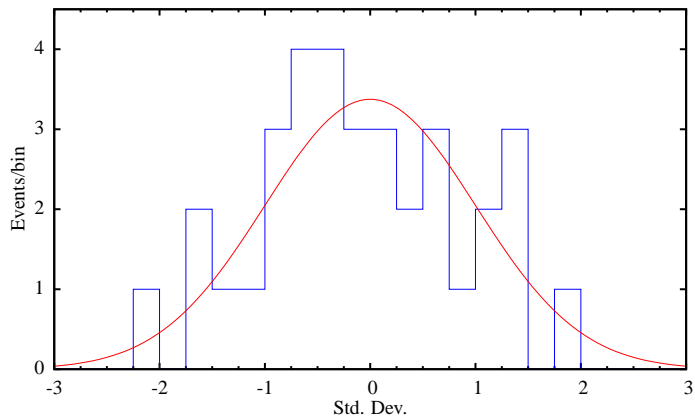}
\caption{Distribution of error residuals in  fit compared to a
  Gaussian distribution. \label{fig:newstats} }
\end{figure}

Before, the distribution of residual errors in the CDF fit did not
have an obvious statistical distribution.  We see in Fig.\
\ref{fig:newstats} that after adding data-derived single-top, the
distribution of errors is a textbook sampling of a Gaussian
distribution.  Hence, we have solved both the normalization issue,
but, more importantly, all shape issues from, 28--300 GeV.  We
consider this extremely strong evidence that the excesses in the CDF
single-top sample and in $Wjj$ have the same origin.  The $Wjj$ excess
is perfectly explained by the shape of single-top.  Whether this
\textit{is} single-top, we'll address more below, but the excess is
$Wjj$ is almost certainly a kinematic shoulder, and not a resonant
particle.

\begin{figure}[htb]
\includegraphics[width=3.75in]{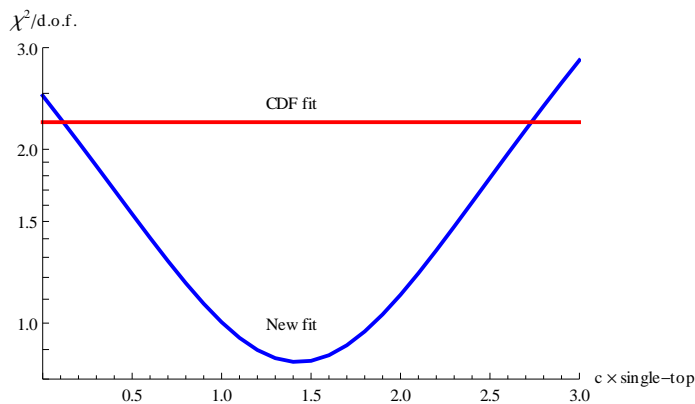}
\caption{Comparison of the $\chi^2/\mathrm{d.o.f.}$ we find from
  fitting the published CDF data (including $1\times$ the Standard
  Model prediction of single-top), and from adding $c\times$ single-top
  as extracted from the CDF data set.\label{fig:chisqfits} }
\end{figure}

In case some are uncomfortable with using $1.4\times$ the central
value of extraction of single-top, we point out this is only
$0.5\sigma$ above the central value.  Nevertheless, we also show in
Fig.\ \ref{fig:chisqfits} that using $1.0\times$ the central value
gives $\chi^2/\mathrm{d.o.f.} = 26.0/26$.  Interestingly, almost any
increase above the baseline Monte Carlo prediction for something with
the shape of single-top dramatically improves the
$\chi^2/\mathrm{d.o.f.}$.

\subsection{Comparison with \DZero}
\label{ssec:dzero}

In our paper \cite{Sullivan:2011hu} we made a prediction that \DZero\
would see (at most) a very small excess in $Wjj$ because their early
single-top data agreed almost perfectly with the Standard Model
prediction.  Since that time, \DZero\ has released a paper
\cite{Abazov:2011af} claiming that their data is consistent with no
excess.  While that is statistically true, \DZero\ 
actually does have a modest excess in $t$-channel production.
Specifically, \DZero\ has found $1.28\times$ $t$-channel, and
$0.94\times$ $s$-channel in Ref.~\cite{Abazov:2011rz}.  Strangely, this
small excess goes in exactly the \textit{opposite} direction as the
CDF excess.  However, we thought it would be useful to use the same
procedure to check our prediction for \DZero.

\begin{figure}[htb]
\includegraphics[width=3.65625in]{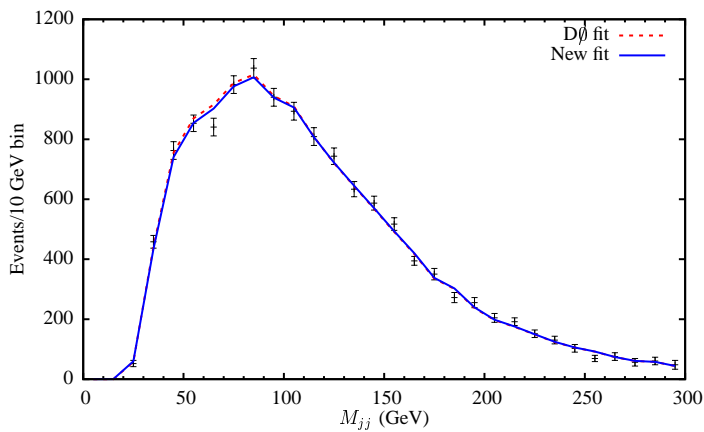}
\caption{Comparison of \DZero\ data with and without additional single-top.
\label{fig:dzero} }
\end{figure}

The result is that the $\chi^2/\mathrm{d.o.f.}$ goes from $26.4/24$ to
$25.3/24$ with no change to the required amount of $WW$ ($b=1$),
slightly less $Wjj_r$ ($a=0.97$), and $1.5\times$ single-top as
extracted from the \DZero\ measurement.  Not much was expected or
needed, but interestingly, the fit is best with almost the same
increase in single-top as CDF.  Obviously this is well within errors.
In Fig.\ \ref{fig:dzero} one can see that above $WW$ threshold adding
a little more single-top is just as consistent as that predicted by
the \DZero\ Monte Carlo.  Hence, this is a suggestive consistency
check, but is not statistically powerful enough to help understand the
CDF anomaly.

\section{Is it single-top?}
\label{sec:what}

We have been very careful to state that the CDF excess in $Wjj$ is
fully explainable by something with the kinematic shape of single-top,
and normalization that fits the criteria of the CDF single-top
measurement excess, but we have been careful not to claim we can prove
it \textit{is} single-top.  The burning question is: what is it?

Initially, we made it clear we were not considering $t\bar t$
production solely because the CDF paper claimed to have fit it in
data, and properly accounted for it.  Given our observation that 3-jet
events feeding into 2-jet events is playing at least some role, this
may need to be revisited.  How well do we understand the sample of
$t\bar t$ under these exact cuts?  We cannot answer that
theoretically.  It requires deep access to the internal procedures
used to fit $t\bar t$ in the first place.

\begin{figure}[htb]
\includegraphics[width=3.65625in]{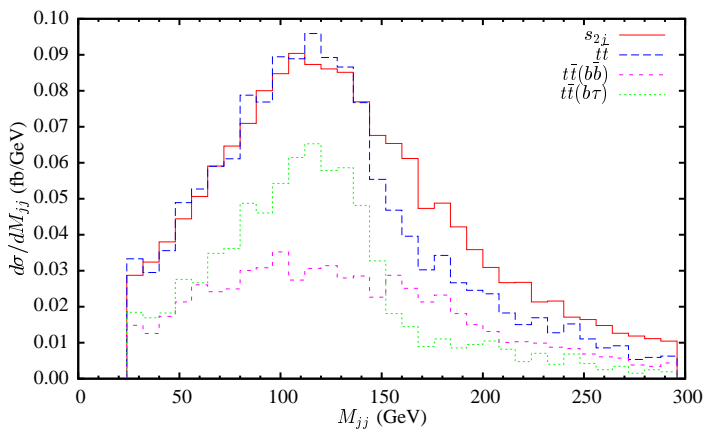}
\caption{$M_{jj}$ for $t\bar t\to b\bar b e\tau$ and $s$-channel single
top into $b\bar be\nu$.\label{fig:tt} }
\end{figure}

Once thing we can show, however, is that the some of the kinematic
shapes in $t\bar t$ are compatible with the single-top shape.  There
are \textit{many} $t\bar t$ final states that could be feeding into
the $Wjj$ analysis.  In Fig.\ \ref{fig:tt} we choose to focus on one:
$t\bar t\to b\bar b e \tau$.  This process could be playing a role, as
$\tau$s are often reconstructed as jets.  Further, they represent only
one particle to miss.  $t\bar t$ certainly has 2 $b$ jets, which an
excess of could help partially explain the anomalously large sample of
2 $b$-tag events in the single-top analysis.

In Fig.\ \ref{fig:tt}, we split out the contributions to $Wjj$ from
$bb$ and $b\tau$ to point out that the shapes are significantly
different.  The shape of $bb$ is almost identical to that of
$s$-channel production (though $t\bar t$ is intrinsically smaller in
this channel, there are other possible decay modes).  $b\tau$ falls
sharply near 150 GeV.  It seems likely that a fit that allowed some
additional $t\bar t\to b\bar b e\tau$ would also fit the data.
However, there are not enough degrees of freedom in the data we have
to trust such a fit.

So where does this leave the question?  The answer is likely that the
excess is due to a combination of top-quark initiated backgrounds plus
something else not understood.  It is in the part that is not
understood that we feel is where the emphasis should be placed.
Having tied together the single-top measurement anomaly and the $Wjj$
excess leaves us with a strange quandary: The CDF $Wjj$ data has too
many $0\, b$-tag events, too many $2\, b$-tag events, and not nearly
enough $1\, b$-tag events in this simple kinematic region.  If
anything, one would expect that a missed $b$-tag would lead to an
excess in the $1\, b$-tag sample at the expense of the $2\, b$-tag
sample.  The size of the effect seems large to be purely statistical,
but that is a detail of the data set that CDF will have to study
internally.  Nevertheless, we feel the solution to this whole
conundrum is tied to explaining this more general discrepancy.

As a final comment on the $b$-tag issue: We had suggested looking for
excess $b$-tags in the $Wjj$ sample early on in this investigation.
The only CDF analysis and plot we have seen claims that $b$'s cannot
be playing a role because the number of $b$ tags between 120--160 is
fractionally the same as outside that region.  Given that a single-top
excess exists across the entire 28--300 GeV region, that is exactly
what we would have predicted.  The question is one of total absolute
rate --- not relative rate in a small mass window.

\begin{figure}[htb]
\includegraphics[width=3.75in]{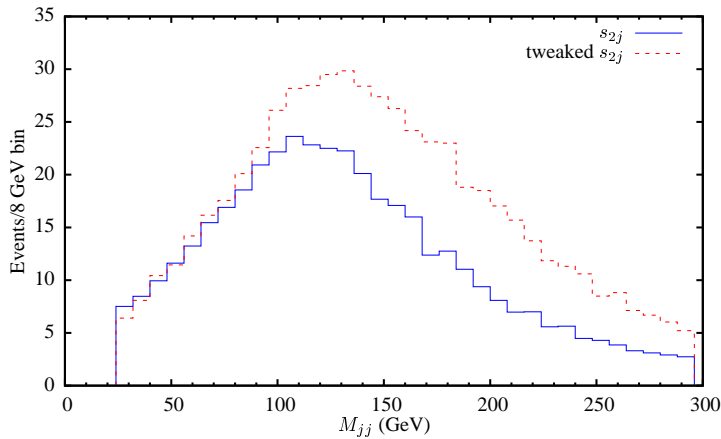}
\caption{Comparison of vanilla LO $s$-channel and $s$-channel distorted by
tweaking measurement uncertainties.  \label{fig:mdd} }
\end{figure}

Given our success in reproducing the Wjj excess without any
alterations to the shape or normalization of single-top (as extracted
from CDF data), it is not necessary to play with jet energy scale
(JES), etc.  However, for completeness, we include a comparison of the
shapes between the MadEvent $s$-channel shape and with what might be
called the ``maximally distorted distribution'' (MDD) in Fig.\
\ref{fig:mdd}.  In the MDD we make the following modifications:
multiply the jet energy scale (JES) by $1.1$, multiply the missing
energy scale by $1.16$, lower the cut on $\MET$ to $\MET > 20$ GeV
from 25~GeV, and lower the cut to $P_{T\,jj} > 20$ GeV from 40~GeV.
The missing energy cut and energy shift are made because they are
sensitive to slightly different physical effects.  The change in the
cut on $P_{T\,jj}$ is made because, as we pointed out in Ref.\
\cite{Sullivan:2011hu} this distribution is not all that well modeled
by leading order.

The conclusions are that shifts in JES, or misprediction on the shape
of $P_{T\,jj}$ can both push the peak of the top backgrounds to
slightly higher $M_{jj}$ values; and allowing uncertainty into the
missing energy measurement can increase the normalization.  Neither is
required, but does give some sense of the inherent uncertainties in
the shapes and normalizations fit in the data.  There are additional
theoretical uncertainties on top of these, but they are a minor issue
compared to understanding the origin of the significant deficit of
$t$-channel production and excess in $s$-channel production in that
3.2 $\fbi$ data set.

In conclusion, if a data-derived background estimation including
single-top-quark production is used to measure $Wjj$ we find that
there is no evidence of any excess or deficit within the CDF data set.
Instead, a more interesting anomaly exists within CDF data: namely,
there are too many $W+0$ and $W+2$ $b$-tag events, and too few $W+1$
$b$-tag events.  Hopefully, an understanding of the origin of this
observation will explain the both the apparent $Wjj$ and
single-top-quark anomalies in the CDF data set.

\begin{acknowledgments}
  This work is supported by the U.~S.\ Department of Energy under
  Contract No.\ DE-FG02-94ER40840.
\end{acknowledgments}

\bigskip 

\end{document}